\documentclass[11pt]{JHEP3}

\usepackage{graphicx}
\usepackage{graphics}
\usepackage{epsfig}

\title{Non-global logarithms in inter-jet energy flow with kt clustering
requirement}

\author{R.B. Appleby and M.H. Seymour \\ Department of Theoretical Physics, Schuster Laboratory,
\\ University of Manchester, UK, M13 9PL \\
E-mail: \email{robert@theory.ph.man.ac.uk}, \email{seymour@theory.ph.man.ac.uk}}

\preprint{MC-TH-2002-11 \\ hep-ph/0211426}


\abstract{Recent work in inter-jet energy flow has identified a class of leading logarithms previously not
considered in the literature. These so-called non-global logarithms have been shown to have significant numerical
impact on gaps-between-jets calculations at the energies of current particle colliders. Here we calculate, at fixed order and to all orders, 
the effect of applying 
clustering to the gluonic
final state responsible for these logarithms for a trivial colour flow 2 jet system. Such a clustering algorithm 
has already been used for experimental measurements at HERA. We find that the impact of the non-global logarithms is 
reduced, but not removed,
when clustering is demanded, a result which is of considerable interest for energy flow observable calculations.}

\begin{document}

\section{Introduction}

The identification of the inter-jet energy flow as an infrared safe way to study gaps-between-jets processes has produced
extensive interest in the last few years. By considering such observables  we may start to formulate a perturbative approach to
the description of such cross sections, as well as probe the interface between perturbative and non-perturbative physics
at energy scales $\sim 1\,GeV$. The inter-jet energy distribution was recently calculated by Sterman {\it et al} 
\cite{Oderda:1998en,Kidonakis:1998nf,Oderda:1999kr,Berger:2001ns} by separating out
the primary emission Bremsstrahlung component and calculating this quantity to all-orders for a 4 jet system. However it was recently pointed out
by Dasgupta and Salam \cite{Dasgupta:2001sh,Dasgupta:2002bw} that this procedure does not include the effects of so-called non-global logarithms, a set of leading logarithms 
which were shown to be numerically important at the energy scales probed by current colliders. 

Recently the H1 and ZEUS collaborations \cite{Adloff:2002em,Zeusdat} performed improved gaps-between-jets analyses in which the entire event
is clustered into (possibly soft) jets using the inclusive kt algorithm \cite{Catani:1993hr,Ellis:tq,Butterworth:2002xg}. A gap event is then defined by the total minijet
energy in the inter-jet region, rather than the total hadronic energy as in previous analyses.
The hope was that this would `clean up' the edges of the gap and make this observable less sensitive to 
hadronic uncertainties. 

In this paper we show that by demanding
the gluonic final state to survive a clustering criterion the effect of these logarithms is reduced, but they are still numerically important,
at HERA and the Tevatron. The study presented here is particularly interesting in the light of the recent analyses by the H1 and ZEUS
collaborations.

The Bremsstrahlung component of the energy flow observable has been the focus of intense work in the last decade, for examples
see  \cite{Berger:2001ns,Contopanagos:1996nh,Catani:1996yz,Kidonakis:1997gm}. It was found 
that by considering the emission of soft, wide angle gluons by light-like quarks one may describe the soft gluon dynamics 
of hadronic processes by an effective theory known as eikonal theory. This subject is now very well developed and the resummation
of such primary logarithms has recently made contact with high-but-fixed order calculations \cite{Sterman:2002qn}. This formalism now allows detailed
calculations of inter-jet energy flow as well as illuminating insights into the topology of colour mixing \cite{Kidonakis:1998nf}.

However, in the study of single-hemisphere observables \cite{Dasgupta:2001sh} Dasgupta and Salam identified that an important class of logarithms
was missing from the soft gluon calculations. These contributions arise in observables that are sensitive to radiation in a restricted
region of phase space and are formally the same order as the primary emission component for inter-jet
observables. To see the origin of these logarithms more clearly, consider soft gluonic radiation
into a patch of phase space $\Omega$ arising from a 2 jet system, where we restrict the total energy of
radiation into $\Omega$ to be less than $Q_{\Omega}$.
The primary logarithms arise from gluons that are emitted directly into $\Omega$ with energy vetoed down to a scale $Q_{\Omega}$ and form the single logarithmic (SL) set 
$(\alpha_s \log \left(\frac{Q}{Q_{\Omega}} \right)^n)$, $n \geq 1$, where $Q$ is the scale of the jet line. Now consider a gluon being emitted outside
of $\Omega$ with intermediate energy $Q_1$, and then vetoing subsequent emission from this gluon into $\Omega$ down to scale $Q_{\Omega}$. Integrating
$Q_1$ up to $Q$ then generates another SL set of $(\alpha_s \log \left(\frac{Q}{Q_{\Omega}} \right)^n)$, $n \geq 2$, 
which are formally the same order
as the primary emission terms. This was studied in detail in \cite{Dasgupta:2002bw,Banfi:2002hw} and in this work we consider how the conclusions 
found are modified phenomenologically when the kt clustering algorithm is applied to the final state.

The organisation of this paper is as follows. Section 2 describes the kt clustering algorithm used
in the H1 and ZEUS analyses, known as the inclusive jet clustering algorithm. We then develop the algorithm, in order to derive a form
suitable for use with a 2-gluon system. Section 3 then describes the order $\alpha_s^2$ calculation of the effect of non-global
logarithms in our 2-jet system, with the clustering algorithm included in the calculation. We find that the asymptotic suppression factor with
clustering is reduced with respect to the non-clustering case. We then proceed to calculate the non-global effect to all orders in section 4 using the combination of the
large $N_c$ limit and a Monte Carlo algorithm and we conclude with a summary and a discussion of further work.

The all-orders treatment used in this work allows direct inclusion of the kt algorithm, in exactly the same way as is used
experimentally, and we find that the clustering process reduces the magnitude of the non-global corrections to the primary suppression factor. However they are still 
phenomenologically relevant at HERA.

\section{The kt clustering procedure}

The version of the
algorithm we use, which is the one used in the HERA analyses, is known as the inclusive kt algorithm \cite{Catani:1993hr,Ellis:tq,Butterworth:2002xg}.
The main features of importance to the present study are: the clustering procedure starts from the particles of lowest relative transverse 
momenta and iteratively merges them to construct pseudoparticles of higher transverse momentum; the decision of whether a particular 
pair of pseudoparticles are merged depends only on their relative opening angle (see below); despite this, it is possible 
for soft particles to be `dragged' through relatively large angles by being merged with harder particles, which are merged with even harder 
particles, and so on. Nevertheless, we do not expect out results to be qualitatively different from other infra-red safe 
jet algorithms, such as the improved Legacy Cone algorithm of \cite{Blazey:2000qt}.
In the remainder of this section we will provide a formulation of the kt algorithm that can be directly applied to our fixed 
order calculation in the next section. We follow the H1 and ZEUS analyses closely and set the radius parameter, $R$, to unity for
most of the numerical results.
The algorithm is implemented in the package KTCLUS \cite{Catani:1993hr}, which is used for the all-orders calculation later in this work.

It is necessary to use the full iterative algorithm for experimental analysis and
for Monte Carlo applications, but when we consider a 2-gluon final state in this work we can reduce the algorithm to
a convenient analytic form. We start by considering a hard jet line at scale $Q$ which radiates a gluon with some transverse 
energy $E_{T,1}$ in some direction $(\eta_1,\phi_1)$. This system then radiates a secondary soft gluon with transverse
energy $E_{T,2}$ in some direction $(\eta_2,\phi_2)$. We assume that the transverse energies of the gluons are strongly ordered, 
\begin{equation}
E_{T,1} \gg E_{T,2}.
\end{equation}
The kt algorithm works by defining for every pair of particles $ij$ a `closeness' $d_{ij}$ and for every particle $i$ a `closeness
to the beam direction' $d_i$. For the 2 gluon final state, these are
\begin{eqnarray}
d_1 &=& E_{T,1}^2, \nonumber \\
d_2 &=& E_{T,2}^2, \nonumber \\
d_{12}&=& E_{T,2}^2 [ (\eta_1 - \eta_2)^2 + (\phi_1 - \phi_2)^2]. 
\end{eqnarray}
By considering the strong ordering of the transverse momenta, the two gluons will be clustered if $d_{ij} < d_2$, so we
require
\begin{equation}
(\eta_1 - \eta_2)^2 + (\phi_1 - \phi_2)^2 > R^2
\end{equation}
for the two gluons to constitute separate jets and not be merged by the algorithm. Therefore for a two-gluon final state to pollute the
gap and generate secondary logarithms we require that gluon 1 is outside the gap, gluon 2 is inside the gap and that they be sufficiently
separated in $(\eta,\phi)$ to avoid being merged. Therefore the clustering condition manifests itself
as a $\Theta$-function in our calculation,
\begin{equation}
\Theta( (\eta_1 - \eta_2 )^2 + \phi_2^2 -R^2 ),
\end{equation}
where we have used our freedom to set $\phi_1=0$. We will use this result in the next section.

\section{Fixed order calculation}

\subsection{Definitions and primary emission form factor}

Following Dasgupta and Salam \cite{Dasgupta:2002bw}, the observable we are interested in is the total transverse energy $E_t$ flowing into
a region of phase space $\Omega$ for an event characterized by the hard scale $Q$,
\begin{equation}
E_t=\sum_{i \in \Omega} E_{t,i}.
\end{equation}
We are specifically interested in the cases of $\Omega$ being either a slice in rapidity or a patch,
bounded in rapidity and azimuthal angle. The quantity we shall calculate is called $\Sigma_{\Omega}$ and is 
defined to be the probability that $E_t$ is less than some energy scale $Q_\Omega$. 
\begin{equation}
\Sigma_{\Omega}=\frac{1}{\sigma_o}\int_0^{Q_{\Omega}} d\,E_t \frac{d\,\sigma}{d\,E_t}.
\end{equation}
We shall assume
the strong ordering $Q_\Omega \ll Q$.
The aim of this work is to calculate the importance of the non-global contribution to 
$\Sigma_{\Omega}$ and so it is convenient to factorize this expression into a function describing 
primary emission into $\Omega$, $\Sigma_{\Omega,P}(t)$, and a function describing (secondary) emission into $\Omega$ from 
large-angle soft gluons outside of $\Omega$, $S(t)$,
\begin{equation}
\Sigma_{\Omega}(t)=S(t)\Sigma_{\Omega,P}(t).
\end{equation}
We have denoted the following integral of $\alpha_s$ by $t$,
\begin{eqnarray}
t(Q_{\Omega},Q)&=&\frac{1}{2\pi}\int_{Q_{\Omega}}^{Q/2} \frac{d\,k_t}{k_t} \alpha_s(k_t), \\
\label{runningt}
&=& \frac{1}{4\pi \beta_0}\log\left(\frac{\alpha_s(Q_{\Omega})}{\alpha_s(Q/2)}\right),  \\
&=& \frac{\alpha_s}{2\pi}\log\frac{Q}{2Q_{\Omega}},
\end{eqnarray}
where the first equality is exact, the second holds at one loop, the third assumes a fixed coupling and $\beta_0=(11 C_A - 2 n_f)/(12\pi)$.
The leading order contribution to $S(t)$ comes in at $\alpha_s^2$ and we shall calculate this for a
2 jet system in the next section but it is useful to first consider the primary emission function at 
first order in $\alpha_s$. If we do not restrict the phase space for gluon emission then, order by order in perturbation 
theory, we expect a complete cancellation of real and virtual soft gluon contributions to the primary emission 
form factor. However the requirement of a gap in a restricted region of phase space results in this cancellation
being spoiled and we are left with an integral over the vetoed region. Hence, to
order $\alpha_s$,
\begin{eqnarray}
\Sigma_{\Omega}^{(1)}(Q_{\Omega},Q)&=&-4C_F\frac{\alpha_s}{2\pi}\int_{Q_{\Omega}}^{Q/2} \frac{d\,k_t}{k_t} \int_{\Omega}
d\,\eta \frac{d\,\phi}{2\pi} \\
&=& -\frac{4C_F\alpha_s}{2\pi} A_{\Omega} \log \left( \frac{Q}{2Q_{\Omega}} \right),
\end{eqnarray}
where $A_{\Omega}$ denotes the area in $(\eta,\phi)$ space of the region $\Omega$.
The assumption that the primary gluons are emitted independently according to a two-particle
antenna pattern means that by exponentiating the one loop answer and running the coupling
to the scale $k_t$, we can write $\Sigma_{\Omega,P}(t)$ to all orders,
\begin{equation}
\Sigma_{\Omega,P}(t)=\exp \left( -4C_F A_{\Omega} t \right).
\end{equation}
This equation only includes contributions from independent primary emission.

\subsection{The function \boldmath$S(t)$}

The non-global contribution to $\Sigma_{\Omega}$ is contained in the function $S(t)$, which has its first
non-trivial term at order $\alpha_s^2$. Hence we can write the following expansion for $S(t)$,
\begin{equation}
S(t)=1+S_2 t^2 + S_3 t^3 + \ldots= 1+\sum_{n=2}S_n t^n.
\end{equation} 
The goal of this section is to calculate $S_2$ for two different geometries of $\Omega$ with the
condition that the topology of the gluon tree satisfies the clustering algorithm defined previously. We begin 
by defining the following 4-momenta,
\begin{eqnarray}
k_a &=& \frac{Q}{2}(1,0,0,-1), \\
k_b &=& \frac{Q}{2}(1,0,0,1), \\
k_1 &=& P_{t,1}(\cosh(\eta_1),\sin(\phi_1),\cos(\phi_1),\sinh(\eta_1)), \\
k_2 &=& P_{t,2}(\cosh(\eta_2),\sin(\phi_2),\cos(\phi_2),\sinh(\eta_2)),
\end{eqnarray}
and for a general region $\Omega$ define $S_2$ by
\begin{eqnarray}
S_2 \log^2\left (\frac{Q}{2 Q_{\Omega}}\right) + \mathcal{O}\left(\log\left(\frac{Q}{2Q_{\Omega}}\right)\right) =
-C_F C_A \int_{k_1 \not\in \Omega}  d\,\eta_1 \frac{d\,\phi_1}{2\pi} \nonumber \\
\int_{k_2 \in \Omega} d\,\eta_2 \frac{d\,\phi_2}{2\pi}
\frac{Q^4}{16} \int_0^1 x_2 d\,x_2 \int_{x_2}^1 x_1 d\,x_1 \Theta\left(x_2-\frac{2Q_{\Omega}}{Q}\right)W_S,
\end{eqnarray}
where we define the transverse momentum fraction by 
\begin{equation}
 P_{t,1} = x_1 \frac{Q}{2}.
\end{equation}
This expression for $S_2$ contains the secondary part, $W_S$, of the well-known matrix element squared for the energy ordered emission
of two gluons, which can be derived from gluon insertion techniques \cite{Dokshitzer:1992ip},
\begin{eqnarray}
W&=&4C_F \frac{(ab)}{(a1)(1b)} \left( \frac{C_A}{2}\frac{(a1)}{(a2)(21)} + \frac{C_A}{2}\frac{(b1)}{(b2)(21)}
+ \left( C_F - \frac{C_A}{2} \right) \frac{(ab)}{(a2)(2b)} \right) \nonumber \\
&=& C_F^2 W_P + C_F C_A W_S,
\end{eqnarray}
where the notation $(ij)$  denotes the dot product of the appropriate 4-momenta. This expression contains the
primary emission piece $W_P$, proportional to $C_F^2$, and the piece that interests us, which is the part
proportional to $C_F C_A$ and denoted $W_S$. Note that the last term is the dipole interference term and 
is absent in the large $N_c$ limit. Denoting the secondary emission piece by $W_2$ and evaluating the 4-momenta 
products we get
\begin{eqnarray}
(ab) &=& \frac{Q^2}{2}, \\
(1a) &=& \frac{Q^2 x_1}{4}\exp(-\eta_1), \\
(1b) &=& \frac{Q^2 x_1}{4}\exp(+\eta_1), \\
(2a) &=& \frac{Q^2 x_2}{4}\exp(-\eta_2), \\
(2b) &=& \frac{Q^2 x_2}{4}\exp(+\eta_2), \\
(12) &=& \frac{Q^2 x_1 x_2}{4}\left( \cosh(\eta_1-\eta_2)-\cos(\phi_1-\phi_2)\right).
\end{eqnarray}
Now setting $\phi_1$ equal to zero we arrive at
\begin{equation}
W_S = \frac{128}{Q^4 x_1^2 x_2^2} \left( \frac{\cosh(\eta_1-\eta_2)}{\cosh(\eta_1-\eta_2)-\cos(\phi_2)}  - 1 \right).
\end{equation}
The energy fraction integrals are straightforward and lead to the following leading logarithm (LL) result
(the $\Theta$-function ensures that sufficient energy reaches the gap region $\Omega$,)
\begin{equation}
\int_0^1 \frac{d\,x_2}{x_2} \int_{x_2}^1 \frac{d\,x_1}{x_1} \Theta\left(x_2 - \frac{2Q_{\Omega}}{Q}\right)=
\frac{1}{2}\log^2\left(\frac{2Q_{\Omega}}{Q}\right)
\end{equation}
Hence we arrive at
\begin{equation}
S_2 =-4 C_F C_A \int_{angles}
\left[ \frac{\cosh(\eta_1-\eta_2)}{\cosh(\eta_1-\eta_2)-\cos(\phi)} -1 \right].
\end{equation}
Specializing to a slice in rapidity of width $\Delta \eta$ and delimited by $|\eta| < \Delta \eta/2$, where
boost invariance allows us to center the gap on $\eta=0$, we can exploit 
the symmetry of the $\eta_1$ integral and
perform the trivial $\phi$ integral of the first gluon to express the final integral over angles as
\begin{equation}
\int_{angles}=2 \int_{-\infty}^{-\frac{\Delta\eta}{2}} 
d\,\eta_1 \int_{-\frac{\Delta\eta}{2}}^{\frac{\Delta\eta}{2}} d\, \eta_2 \int_0^{2\pi} \frac{d\,\phi_2}{2\pi}.
\end{equation}
For the case of no clustering of the gluons it is possible to perform the azimuthal average using the result,
\begin{equation}
\int_0^{2\pi} \frac{d\, \phi}{2\pi} \frac{1}{\cosh(\eta_1-\eta_2)-\cos(\phi_2)} = \frac{1}{|\sinh(\eta_1-\eta_2)|},
\end{equation}
which can be proved by noting that the following contour integration over the unit circle,
\begin{equation}
I=\frac{2}{i}\oint_{u_1} \frac{d\,z}{(z-\exp(\eta_2-\eta_1))(z-\exp(\eta_1-\eta_2))},
\end{equation}
is equivalent to the integration over the azimuthal angle of the second gluon.
For the case of no clustering of the gluons, it is possible to perform all of the integrations analytically and obtain \cite{Dasgupta:2002bw},
\begin{equation}
S_2 = -4C_FC_A \left[ \frac{\pi^2}{12} + (\Delta\eta)^2 - \Delta\eta \log(e^{2\Delta\eta}-1) - \frac{1}{2}\mathrm{Li}_2 (e^{-2\Delta\eta}) -  \frac{1}{2}\mathrm{Li}_2 (1-e^{2\Delta\eta})\right].
\end{equation}
Note that as $\Delta\eta$ increases, $S_2$ rapidly saturates at its asymptotic value,
\begin{equation}
\lim_{\Delta\eta \rightarrow \infty}=-C_FC_A\frac{2 \pi^2}{3}.
\end{equation}
This analytic evaluation of $S_2$ is not possible for the case of clustered gluons because, as discussed 
previously, the requirement that a gluonic final state is in a configuration that will survive a clustering
algorithm can be written as a $\Theta$-function of all three angular integration variables,
\begin{equation}
\Theta( (\eta_1 - \eta_2 )^2 + \phi_2^2 -R^2 ).
\end{equation}
However we can readily reduce the three-dimensional integral to a one-dimensional integral using standard
techniques if we consider the region of phase space that is vetoed by the clustering algorithm, denoted $S_2^v$. Doing this 
we obtain,
\begin{equation}
S_2^v=\frac{-32 C_F C_A}{\pi} \int_0^R \mathrm{min}(\eta,\Delta\eta)\bigg[ 2 \coth(\eta) \arctan\left(\frac{\tan(\sqrt{R^2-\eta^2}/2)}{\tanh(\eta/2)}
\right) \nonumber - \sqrt{R^2-\eta^2} \bigg] d\,\eta,
\end{equation}
where we define $\eta=\eta_1-\eta_2$.
Therefore the solution for $S_2$ with the clustering condition imposed is obtained by subtracting $S_2^v$ from the analytic unclustered
result.
We resort to  numerical techniques to solve this equation, which has the advantage of being easily 
extendible to any gap geometry of interest. Our numerics were done using Monte Carlo integration methods
and figure \ref{figa} shows our result for $S_2$ as a function of $\Delta\eta$ for a cluster radius of
$R=1.0$.
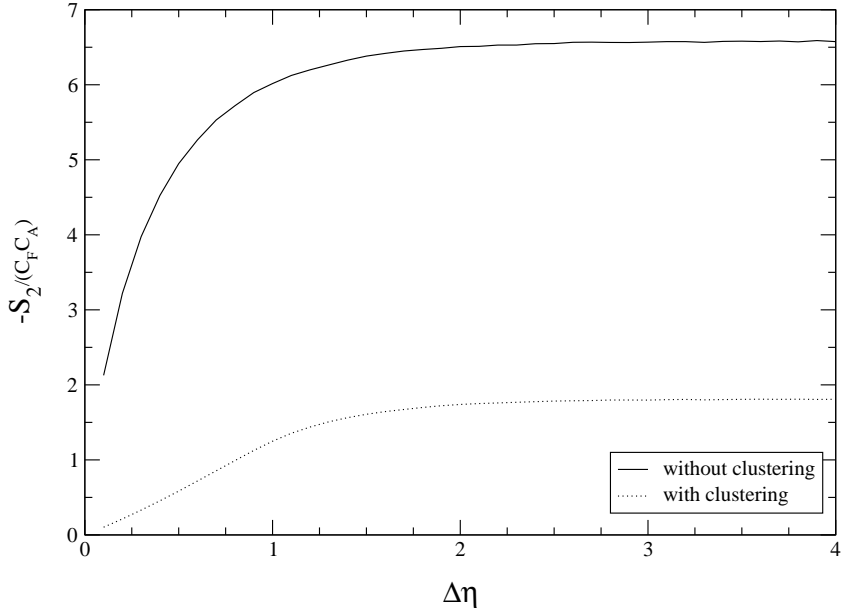
\FIGURE[p]{\epsfig{file=a.eps}\caption{$S_2$ as a function of $\Delta\eta$. The clustered curve is obtained using a kt algorithm with
a radius parameter of 1.0.}\label{figa}}
In this figure we also show, as the solid line, the curve for $S_2$ which is obtained with
no clustering condition imposed on the gluons. In \cite{Dasgupta:2002bw} the saturation for $S_2$ was explained by 
observing that the dominant contribution to $S_2$ arises from the situation with one gluon just outside the  gap emitting a gluon just
inside the gap. This occurs when $\eta_1 \simeq \eta_2 \simeq -\Delta\eta/2$ and the dominant contribution arises from the collinear
region of the matrix element. The results we have obtained show that when we demand the gluonic final state to survive a clustering
algorithm, the saturation of $S_2$ observed in \cite{Dasgupta:2002bw} is still observed but the saturation value is reduced by $70\%$. In other words
the value that $S_2$ saturates to is reduced from $6.57$ to $1.81$.
The reason for this reduction is that we have removed gluons from the region of collinear enhancement but gluons that are still
sufficiently separated in the $(\eta,\phi)$ plane to satisfy
\begin{equation}
(\eta_1 - \eta_2 )^2 + \phi_2^2 > R^2
\end{equation}
will survive the clustering process and contribute to the non-global logarithms.
Therefore we conclude that the saturation of the non-global contribution at fixed order is still seen when we demand clustering 
of the final state. If this reduction holds to all orders then the magnitude of the non-global logarithms will be smaller for the clustered
case relative to the non-clustered case. To explore this it is necessary to perform the all-orders treatment, which we do in the next section.

\section{All-orders calculation}

The extension of the above calculation to all orders presents considerable mathematical problems due to the geometry and
colour structure of the gluon ensemble. Therefore we have used the large $N_c$ approximation and numerical methods, as
developed by Dasgupta and Salam \cite{Dasgupta:2002bw}, to extend 
the calculation to all orders. We have extended the simulation developed in the work
of Dasgupta and Salam
to include clustering of gluons according to the kt algorithm discussed previously. This is
performed by terminating the gluon cascade evolution if $\Omega$ is polluted and the final
state survives the clustering algorithm. If the latter condition is violated, the Monte Carlo algorithm 
is allowed to continue.

Note that this algorithm can produce configurations with
many gluons at high $t$  and this, coupled with the fact that the speed of the kt algorithm scales 
like $N^3$ (where $N$ is the number of gluons), means that the Monte Carlo algorithm can be very slow in this region.
The result is that we have large statistical errors for $t \ge 0.3$, so we do not show this region.
In the next section we shall present our results.

\subsection{All-orders results}
We have verified the results obtained by Dasgupta and Salam \cite{Dasgupta:2002bw}, which have been calculated with no clustering requirement imposed on the gluons.
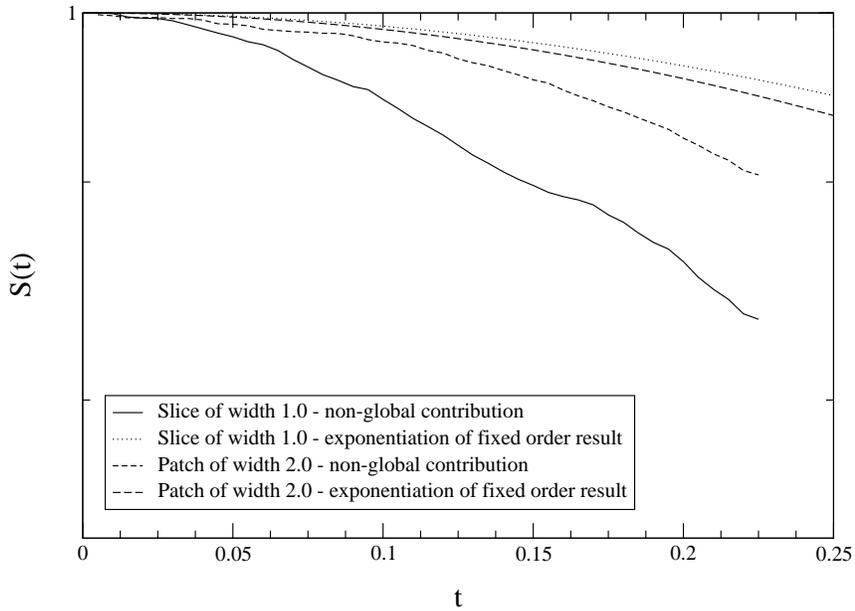
\FIGURE[p]{\epsfig{file=b.eps}\caption{The function $S(t)$ for a slice and a patch of phase space, with the condition of kt clustering imposed on the gluons. These curves
were obtained using a Monte Carlo procedure in the large $N_c$ limit. Also shown are the curves for $S(t)$ obtained by exponentiating
the fixed order result. The geometry independence of the $t$ dependence indicates that the buffer zone mechanism identified to exist in previous work
survives the clustering algorithm.}\label{figb}}

Figure \ref{figb} shows the function $S(t)$ for two different geometries for $\Omega$: a slice in rapidity 
of width $\Delta\eta=1.0$ and a square patch in rapidity and azimuthal angle of side length $\Delta\eta=\Delta\phi=2.0$,
with the requirement of clustering on the final state gluons. This is implemented in the Monte Carlo algorithm
in the same way as it is used by H1 and ZEUS, allowing us to estimate the experimental impact of the clustering scheme on the non-global 
logarithms. Firstly the figure shows that at high $t$ ($t > 0.2$) the suppression increases faster for the full calculation than for the
exponentiation of $S_2$ with clustering and the two curves have different shapes in this region. This agrees with the analysis of
the case with no clustering. Secondly we see that the $t$ dependence of the suppression
is geometry independent at high $t$; the implication of this is that the clustering process maintains a buffer region\footnote{The buffer region is defined
by the absence of any reconstructed jets within it. It may therefore contain gluons, provided that they get pulled
out of the buffer region by the clustering algorithm.} of suppressed intermediate
radiation around $\Omega$ exactly as in the unclustered case. Therefore figure \ref{figb} shows that the clustering requirement does not change the gross
features of the non-global suppression at all orders and the results indicate that the buffer region postulated to exist around the gap $\Omega$ is
preserved by the clustering process.

Figure \ref{figc} shows the reduction of the phenomenological significance of the non-global logarithms when we cluster the final state. The figure shows
the function $\Sigma(t)$ with only primary emissions and the full all-orders treatment with and without clustering effects. 
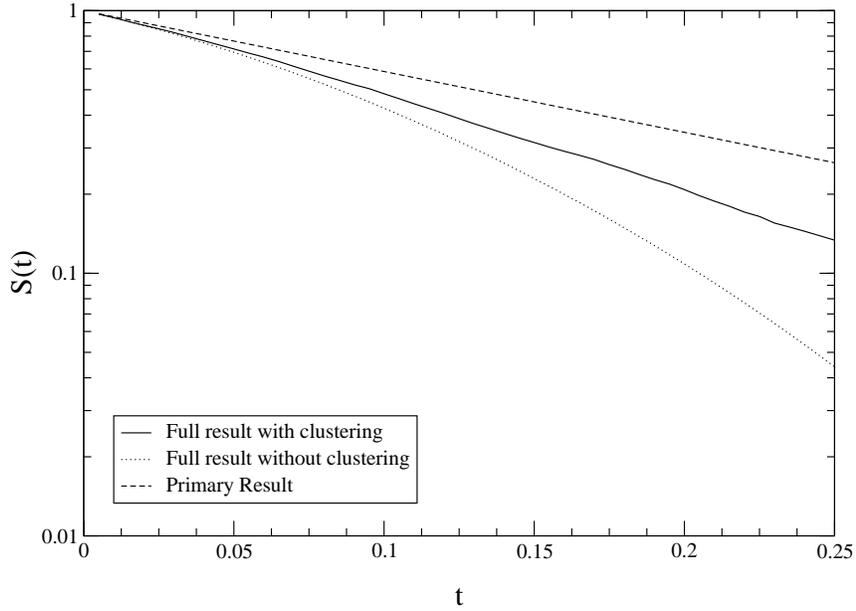
\FIGURE[p]{\epsfig{file=c.eps}\caption{The phenomenological implications of clustering for the case of $\Omega$ being a slice of rapidity of width $1.0$. The reduction
of the non-global suppression factor when clustering is included can be seen.}\label{figc}}
This is done with 
$\Omega$ defined as a slice in rapidity of size $\Delta\eta = 1.0$. There are several points to note. Firstly the effect of the non-global logarithms is a large 
suppression of the cross section relative to the primary-only result. The value of $t$ which we can consider the realistic upper limit for next 
generation experiments
is about $0.15$ so we shall take $t=0.15$ as our reference value. We can translate this into a energy scale by using the running coupling definition 
of t, equation \ref{runningt}, and obtain $Q\sim 100\,GeV$ for $Q_{\Omega}=1\,GeV$. In this region the inclusion of the non-global effects without clustering increases 
the suppression relative to the primary-only result by about a factor of $1.65$. When we include clustering effects, this difference
is reduced to about $1.2$. Therefore, at all orders, the requirement of clustering on the final state reduces the phenomenological significance of the
non-global effects by about 70\%. This reduction in magnitude of the effect can be seen to persist to all orders. Hence when calculating cross sections, if we exclude the effect of non-global logarithms then we will overestimate the cross section by $65\%$ for a non-clustered final state and by $20\%$ for
a clustered final state. At larger $t$ values, the overestimation increases. For comparison, the typical errors on the H1 gaps-between-jets data is
$\sim 30\%$.
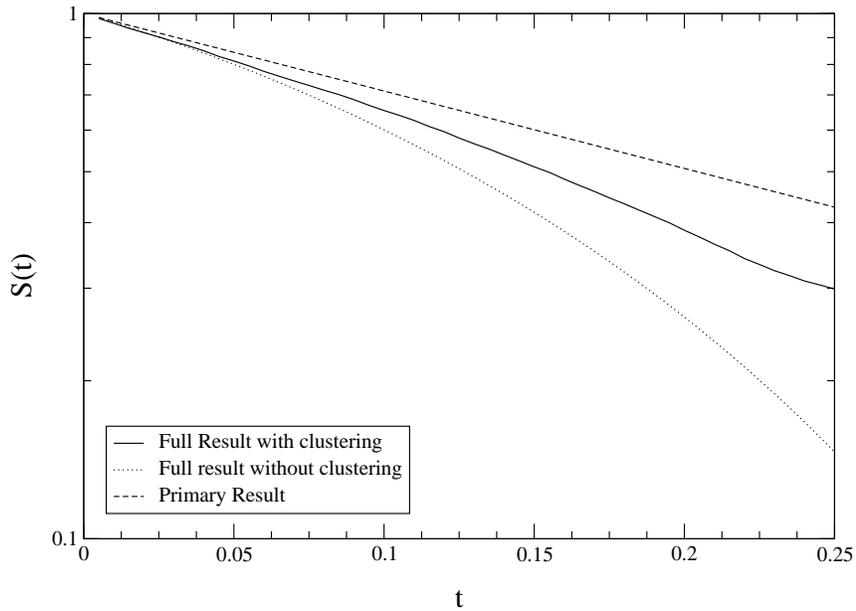
\FIGURE[p]{\epsfig{file=d.eps}\caption{The effect of the clustering of the gluons on the function $S(t)$ for a patch of phase space of size  $\Delta\eta = \Delta\phi = 2.0$}\label{figd}}

Finally, figure \ref{figd} shows the same comparison of the full result with and without clustering for a patch 
of size $\Delta\eta = \Delta\phi = 2.0$. We can see
that the conclusions we made for figure \ref{figc} apply and the effect of the non-global logarithms is of similar magnitude.

We have also performed calculations to see how the non-global suppression is affected by varying the radius parameter, $R$, 
of the clustering procedure.
By decreasing $R$ the impact of the clustering algorithm is reduced and the effect of the non-global logarithms is restored
to the non-clustered case.
We expect the magnitude of the non-global suppression to tend to the non-clustered case as $R \rightarrow 0$.
Similarly, increasing the radius causes more gluons to be included in the clustering and hence 
the magnitude of the suppression to reduce. In fact, we have found that for $R$ close to $1.5$ the full result is almost
identical to the primary result.  

In summary, non-global logarithms are important not only from the point of view of correctness of the leading logarithm series but also result in significant
numerical corrections to cross sections. These corrections are reduced by about a factor of 3 if we cluster the final state. 
It is clear, therefore, that while the use of the kt algorithm has aided the control of the non-global logarithms, they still have a 
significant numerical effect at HERA.

\section{Conclusions}
The observation that observables that are sensitive to radiation in only a restricted part of phase space, so-called non-global 
observables, are strongly sensitive to secondary radiation is a new and exciting discovery. For a long time
it was widely thought, now it seems incorrectly, that it was sufficient to only consider primary emission contributions to such observables.
These primary, or Bremsstrahlung, contributions are well understood for a variety of processes. Recent measurements of gaps-between jets 
events at H1 and ZEUS are exactly the class of observable that are sensitive to these effects and to deal with this fact, a study of
non-global logarithms in the context of HERA measurements is required.
In this work we consider the non-global logarithms generated from secondary radiation into a restricted region of phase space under the condition of 
the final state surviving a clustering algorithm. Such kt clustering algorithms have been used for the HERA measurements. In this paper we have extended
 the study of Dasgupta and Salam and found, for a 2-jet system with clustering imposed, that the clustering process
reduced, but did not eliminate, the non-global logarithm effect.

Our main conclusion is that the final state specified in the H1 and ZEUS analysis will mean that a primary emission calculation in the manner
of Sterman {\it et al} \cite{Berger:2001ns,Oderda:1999kr,Oderda:1998en,Kidonakis:1998nf} of energy flow observables will overestimate the observed gaps-between-jets rate by around 20\%. This is to be contrasted with an overestimation of 65\%
that would be found for a non-clustered final state.
This result was calculated to all orders in the large $N_c$ limit for a 2 jet system. We reserve the extension of these
results to the more complex 4 jet case for future work but we expect the gross conclusions of this study to apply.

\section*{Acknowledgments}
We would like to acknowledge M. Dasgupta for inspiring discussions and for useful comments on the manuscript and to 
G. Salam for supplying the numerical code used in this work.

\end{document}